\documentclass{pas}

\usepackage{multirow}
\usepackage{graphicx}
\usepackage{hyperref}
\usepackage{textcomp}
\usepackage{amsmath}
\usepackage{pifont}
\usepackage{orcidlink}
\usepackage{placeins}

\usepackage[
  backend=biber,
  style=pasa-td,         % your custom biblatex style (.bbx)
  citestyle=authoryear,  % author-year citations
  giveninits=true,
  maxcitenames=2,
  mincitenames=1,
  maxbibnames=8,
  minbibnames=3,
  uniquename=false,
  uniquelist=false,
  doi=false, url=false, isbn=false, eprint=false,
  sorting=nyt
]{biblatex}

\AtEveryCite{%
  \DeclareDelimFormat{finalnamedelim}{\addspace\&\space}%
  \DeclareDelimFormat{multinamedelim}{\addcomma\space}%
}

\DefineBibliographyStrings{english}{andothers = {et\ al\adddot}}

\DeclareFieldFormat{citetitle}{#1}
\DeclareFieldFormat{labeltitle}{#1}

\begin{document}

\lefttitle{Publications of the Astronomical Society of Australia}
\righttitle{Dunn et al.}

\jnlPage{1}{4}
\jnlDoiYr{2021}
\doival{10.1017/pasa.xxxx.xx}

\articletitt{Research Letter}

\title{A century of change: new changing-look event in Mrk 1018's past}

\author{\href{https://orcid.org/0009-0006-7608-1275}{\gn{Thomas} \sn{Dunn}}$^{1,2,*}$, 
\href{https://orcid.org/0000-0002-7960-5808}{\gn{Rebecca} \sn{McElroy}}$^{1,2,3,4}$,
\gn{Mirko} \sn{Krumpe}$^{5}$,
\href{https://orcid.org/0000-0003-2880-9197}{\gn{Scott M.} \sn{Croom}}$^{4}$,
\href{https://orcid.org/0000-0003-2754-9258}{\gn{Massimo} \sn{Gaspari}}$^{6}$,
\href{https://orcid.org/0000-0001-5654-0266}{\gn{Miguel} \sn{Perez-Torres}}$^{6}$,
\href{https://orcid.org/0000-0002-4653-8637}{\gn{Michael} \sn{Cowley}}$^{7}$,
\href{https://orcid.org/0000-0002-3649-5362}{\gn{Osase} \sn{Omoruyi}}$^{9}$,
\href{https://orcid.org/0000-0002-5445-5401}{\gn{Grant} \sn{Tremblay}}$^{9}$,
\href{https://orcid.org/0000-0001-5687-1516}{\gn{Mainak} \sn{Singha}}$^{10,11,12}$
}
\affil{$^1$Centre for Astrophysics, University of Southern Queensland, Toowoomba, QLD 4350, Australia \\
$^2$School of Mathematics and Physics, Physics Annexe, The University of Queensland, St Lucia, Brisbane, QLD 4072, Australia \\
$^3$School of Mathematical \& Physical Sciences, Macquarie University, 12 Wally’s Walk, Macquarie Park, NSW 2113, Australia \\
$^4$Sydney Institute for Astronomy, School of Physics, Physics Road, University of Sydney, NSW 2006, Australia  \\
$^5$Leibniz-Institut für Astrophysik (AIP), An der Sternwarte 16
14482 Potsdam, Germany \\
$^6$Department of Physics, Informatics and Mathematics, University of Modena and Reggio Emilia, 41125 Modena, Italy \\
$^7$Instituto de Astrofísica de Andalucía, Gta. de la Astronomía, Genil, 18008 Granada, Spain \\
$^8$School of Physics and Chemistry, Science and Engineering Centre, Gardens Point campus, 2 George Street, Brisbane, QLD 4000, Australia \\
$^9$Center for Astrophysics $|$ Harvard \& Smithsonian, 60 Garden St., Cambridge, MA 02138, USA\\
$^{10}$Astrophysics Science Division, NASA, Goddard Space Flight Center, Greenbelt, MD 20771, USA\\
$^{11}$Department of Physics, The Catholic University of America, Washington, DC 20064, USA\\
$^{12}$ Center for Research and Exploration in Space Science and Technology, NASA Goddard Space Flight Center, Greenbelt, MD 20771, USA\\
$^*$ Corresponding author, Email: \href{mailto:t.m.dunn@uq.net.au}{t.m.dunn@uq.net.au}
}

\corresp{T. Dunn, Email: \href{mailto:t.m.dunn@uq.net.au}{t.m.dunn@uq.net.au}}

\citeauth{Dunn et al., {\it Publications of the Astronomical Society of Australia}}

\history{(Received 21 Oct 2025; revised xx xx xxxx; accepted xx xx xxxx)}

\begin{abstract}
We investigate the long-term variability of the known Changing Look Active Galactic Nuclei (CL AGN) Mrk 1018, whose second change we discovered as part of the Close AGN Reference Survey (\href{https://cars-survey.github.io/}{CARS}). Collating over a hundred years worth of photometry from scanned photographic plates and five modern surveys we find a historic outburst between $\sim$1935--1960, with variation in Johnson B magnitude of $\sim$0.8 that is consistent with Mrk 1018's brightness before and after its latest changing look event in the early 2010s. Using the combined modern and historic data, a Generalised Lomb-Scargle suggests broad feature with $P=29-47$ years. Its width and stability across tests, as well as the turn-on speed and bright phase duration of the historic event suggests a timescale associated with long-term modulation, such as via rapid flickering in the accretion rate caused by the Chaotic Cold Accretion model rather than a strictly periodic CL mechanism driving changes in Mrk 1018. We also use the modern photometry to constrain Mrk 1018's latest turn-off duration to less than $\sim$1.9 years, providing further support for a CL mechanism with rapid transition timescales, such as a changing mode of accretion.

\end{abstract}


\maketitle

\section{Introduction}
Changing Look Active Galactic Nuclei (CL AGN) are objects observed to undergo dramatic transitions between bright broad-line AGN (Type 1) to dimmer continuum, narrow-line (Type 2) spectral states on timescales of years (\cite{osterbrock_ngc_1976, temple_bass_2022, ricci_changing-look_2023}). With these significant changes occurring on human time spans, studying these objects is an important method of probing accretion physics and active galactic nuclei. 

Mrk 1018 is a nearby late-stage merger at $ z= 0.043$ and is one of the only AGN observed to change spectral type twice since first identified (\cite{mcelroy_close_2016}). This behaviour has been the focus of intensive study (e.g., \cite{krumpe_close_2017, saha_close_2025}). First categorised as a Seyfert 1.9 in 1981 (\cite{osterbrock_seyfert_1981}), five years later when re-observed, the broad lines had brightened sufficiently to reclassify it as a type 1 (\cite{cohen_variability_1986}). Then between 2011--2016, Mrk 1018’s luminosity rapidly declined and spectroscopic observations showed a return to a type 1.9 state (\cite{mcelroy_close_2016}). Since then, the galaxy has been under close investigation, with a notable bright outburst seen in 2020 (\cite{brogan_still_2023}).

Several mechanisms have been proposed to explain the changing look phenomena. Terms to separate two distinct groups were coined by \cite{ricci_changing-look_2023}, Changing-Obscuration (CO) events originate from intervening material, versus the Changing-State (CS) behaviour due to some intrinsic effect altering the amount of accretion onto the central SMBH. Although Mrk 1018 belongs to the latter group (\cite{mcelroy_close_2016, husemann_close_2016}), the mechanism that caused the two transitions has been much discussed (\cite{husemann_close_2016, feng_magnetic_2021, veronese_interpreting_2024}). 
A change in disk accretion mode is a popular (\cite{lyu_long-term_2021, li_interplay_2024, veronese_interpreting_2024, dong_newly_2025, saha_close_2025, adhikari_changing-look_2025}) theory behind Mrk 1018's changing state as first described by \textcite{noda_explaining_2018}. In this model the accretion disk switches between stable states of differing accretion rates causing the changes in continuum strength and line emission. \textcite{veronese_interpreting_2024} propose significant brightening events, such as the one in 2020 (\cite{brogan_still_2023}) could be evidence for an Advective Dominated Accretion Flow (ADAF, one of the modes of accretion engaged by this mechanism), and that long-term monitoring of Mrk 1018's faint state may reveal more. They also favour the alternative, complementary view of Chaotic Cold Accretion (CCA; e.g., \cite{gaspari_chaotic_2013}) driving the changing accretion modes. In this framework, turbulent, thermally unstable halos condense multiphase clouds that collide, cancel angular momentum, and intermittently boost the inflow. CCA naturally yields bursty, stochastic AGN feeding with decade-scale correlation times and a flicker-noise power spectrum (with bends near the characteristic feeding/feedback timescale; \cite{gaspari_linking_2020}), offering a testable explanation for Mrk 1018’s CL cycles without requiring strict periodicity.

Other proposed mechanisms can be tested through their periodic nature, such as a binary black hole (BBH) system (\cite{husemann_close_2016}) or the highly elliptical orbit of a recoiling SMBH (rSMBH), the result of a BBH merger (\cite{kim_recoiling_2018}). Although there is evidence against both mechanisms in the case of Mrk 1018 (\cite{hutsemekers_spectropolarimetry_2020, walsh_vlbi_2023}), if any periodic mechanism is behind its Changing Look, it is expected to repeat every 28--34 years based off the initial change between 1978 and 1984 and the secondary occurring around 2012. Hence, future (or past) changes in spectral type or lack thereof would provide additional evidence for or against such mechanisms.

With recent plate digitisation campaigns, we can look back at Mrk 1018's history to search for major changes in the past, providing insight into the mechanisms behind its Changing Look nature. In this letter we use photographic plates digitised by Harvard College Observatory's Digital Access to a Sky Century at Harvard (DASCH; \cite{williams_dasch_2025}), we build a comprehensive light curve of Mrk 1018's historic brightness over the past hundred years. 

\hyperref[sec:historic]{Section 2} describes the historical data, \hyperref[sec:modern]{Section 3} the modern photometry, \hyperref[sec:analysis]{Section 4} our analysis, \hyperref[sec:discussion]{Section 5} the discussion and \hyperref[sec:conclusion]{Section 6} the conclusions.

\section{Historic Data}
\label{sec:historic}
Over 400,000 photographic plates have been scanned and made available by the Harvard College Observatory's DASCH team (\cite{williams_dasch_2025}). Each plate scan has astrometric accuracy of 1--5 arcseconds and photometric accuracy of 0.1 magnitudes, calibrated to the Johnson B band. Measurements are anchored to the Hubble Guide Star Catalogue and AAVSO Photometric All-Sky Survey (APASS) catalogue to match the 4--17 mag range of the plate collection (\cite{los_progress_2012}). The DASCH database represents a new window into the past for transient and time domain astronomy. Using their custom Python package \textit{daschlab} to interact with the catalogue, we developed a pipeline to filter and display the data alongside modern photometry. 

After source extraction, DASCH flags potentially unreliable detections using standard AFLAG criteria. For Mrk 1018, however, we found these flags often over-rejected detections. We therefore manually inspected all candidate scanned plate images to ensure data quality and detect when Mrk 1018 was resolved. Plate quality being the significant factor, a resolved detection by-eye was considered when Mrk 1018 was discernible above the background noise. Of the 8572 plates covering Mrk 1018, 160 yielded such resolved detections. Due to an offset in position observed on some plate scans (Mrk 1018 was resolved, but not at the correct coordinates), an additional astrometric cut of 15'' was employed to remove these scans. This left 106 plate scans with confirmed sources at Mrk 1018's position for analysis.

DASCH uses an isophotal photometry method whereby source magnitudes are derived from contiguous pixels above a threshold magnitude corresponding to background noise (\cite{laycock_digital_2010}). With this method, an AGN like Mrk 1018 will have a composite magnitude corresponding to host flux and AGN flux.

Isophotal magnitudes may include flux leakage from nearby stars if the sources are blended, as is the case for many dense star regions on saturated plates. These blended sources are detected and flagged during DASCH processing with a cross-check of their reference catalogue for more than one star within half of a detected source's full width half maximum radius. If multiple sources are within this distance, they are flagged as blended objects (\cite{tang_improved_2013}). Out of the 106 visually inspected plate scans, one was flagged as blended this way and removed from processing. Through this method, the remaining sources do not contain flux leakage from the nearby companion stars.

\begin{figure}
    \centering
    \includegraphics[width=0.65\linewidth]{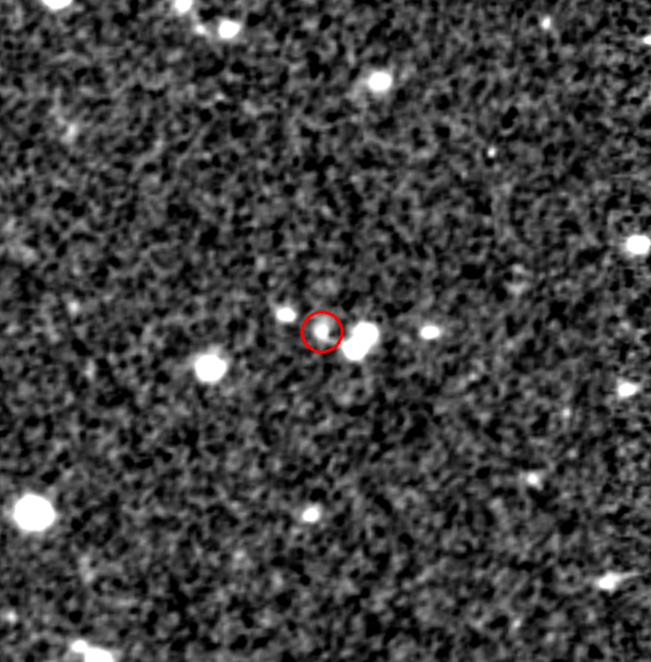}
    \caption{Example plate scan taken from 1935 by the 3-inch Ross Fecker telescope. Despite the small aperture, Mrk 1018 is clearly detected (red circle). This plate was exposed for 8160s, slightly longer than the average of 5847s for the other plates with resolved detections taken by the same telescope.}
    \label{fig:1935_plate}
\end{figure}

These photographic plates come from 15 telescopes ranging from 1 to 24 inches in size. Although some instruments were very small, their photometry is consistent with the larger telescopes, as seen in later analysis. Fig. \ref{fig:1935_plate} shows an example plate scan cutout with a resolved detection of Mrk 1018 in the centre taken from one of the two 3-inch Ross Fecker telescopes whose plates were digitised by DASCH. Both Ross Fecker telescopes (one in Bloemfontein South Africa, and the other Cambridge then Oak Ridge, Massachusetts) provide the most detections as seen in Figs. \ref{fig:mrk1018_B} \& \ref{fig:historic_only} below, despite their small size.

\begin{figure}[!t]
    \centering
    \includegraphics[width=1\linewidth]{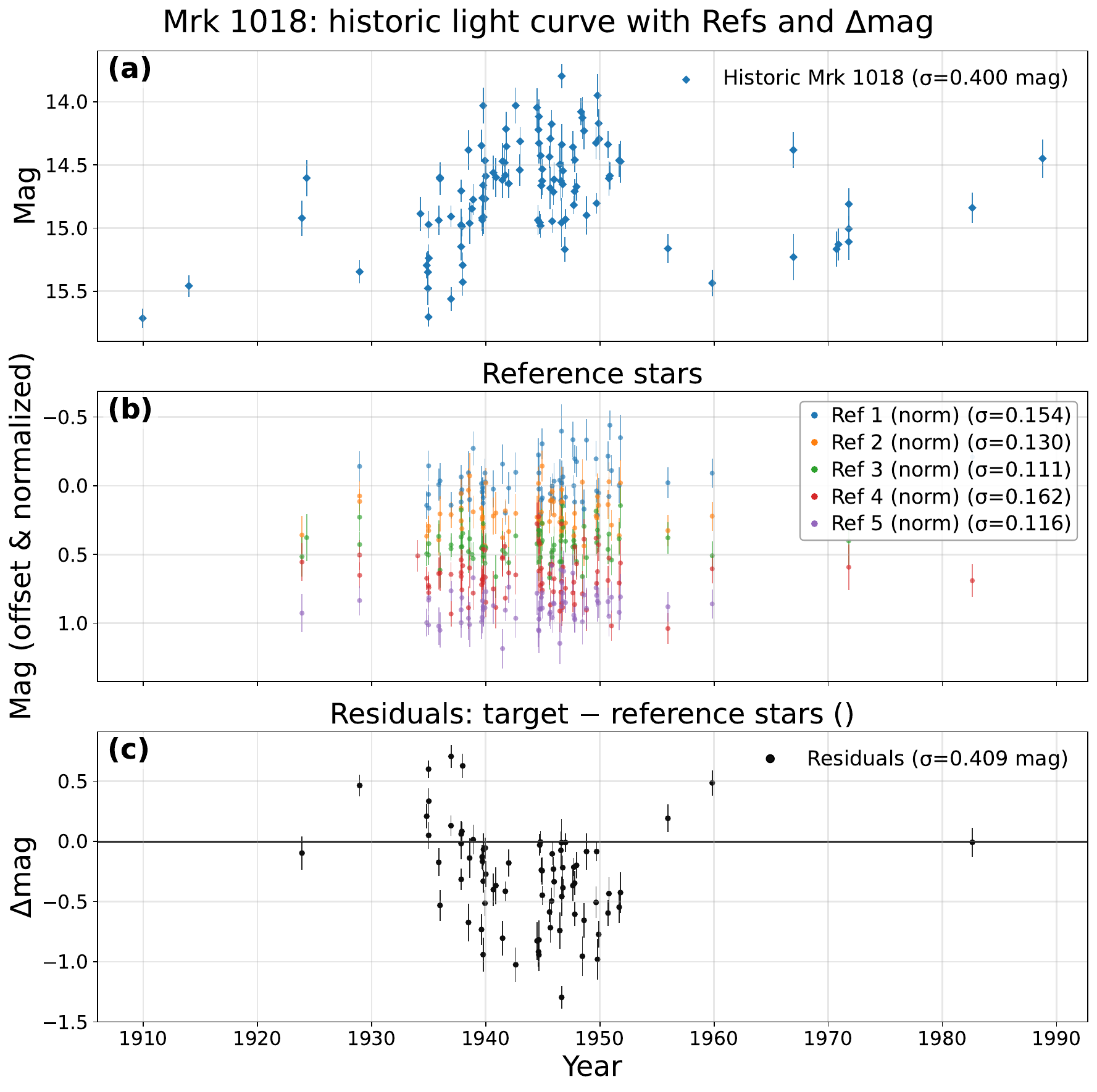}
    \caption{Subplot a: Mrk 1018 AGN + host galaxy historic data (in Johnson B band magnitude). Subplot b: five nearby, similarly bright and coloured reference stars. Subplot c: residuals of the mean magnitude of the reference stars at each epoch (weighted by error) subtracted from Mrk 1018 magnitude. This shows the variability in Mrk 1018 is not associated with plate systematics.}
    \label{fig:mrk1018_with_refs}
\end{figure}

Next, we attempt to verify if the trends seen in the light curve of Mrk 1018 are real by comparing to same-plate detections of nearby stars of similar brightness ($\pm\ 1$ mag) and $B-V$ colour ($\pm\ 0.5$) as per the selected APASS DR8 reference catalogue used by DASCH. Sources of a similar median FWHM ($\pm\ 25\%$ of Mrk 1018's median FWHM) from each of the plate scans. Hence, these reference stars are selected within the ranges, $14.66\pm\ 1$ magnitudes, $B-V=0.99\pm0.5$, and $FWHM=0.01\pm\ 0.003''$. These stars are filtered through DASCH's standard AFLAGs to reject any flawed detections such as those impacted by poor plate quality or a high background level. Despite the AFLAGs over rejecting in Mrk 1018's case, we use reference stars with enough accepted detections during the important period of most variability, 1930--1950 (a 50\% threshold ensures enough reference stars). The five of these reference stars with the most detections can be seen in Fig. \ref{fig:mrk1018_with_refs} subplot b. No intrinsic variability cut is performed, instead the residual is calculated from the median magnitude of these reference stars per plate date subtracted from Mrk 1018's magnitude. This is shown in subplot c of Fig. \ref{fig:mrk1018_with_refs}. From these reference star brightness trends and their residuals against Mrk 1018's brightness, we see no systematical correlations with specific photographic plates for sources of similar appearance to Mrk 1018.

\section{Modern Data}
\label{sec:modern}

\begin{figure}[h]
    \centering
    \includegraphics[width=\textwidth]{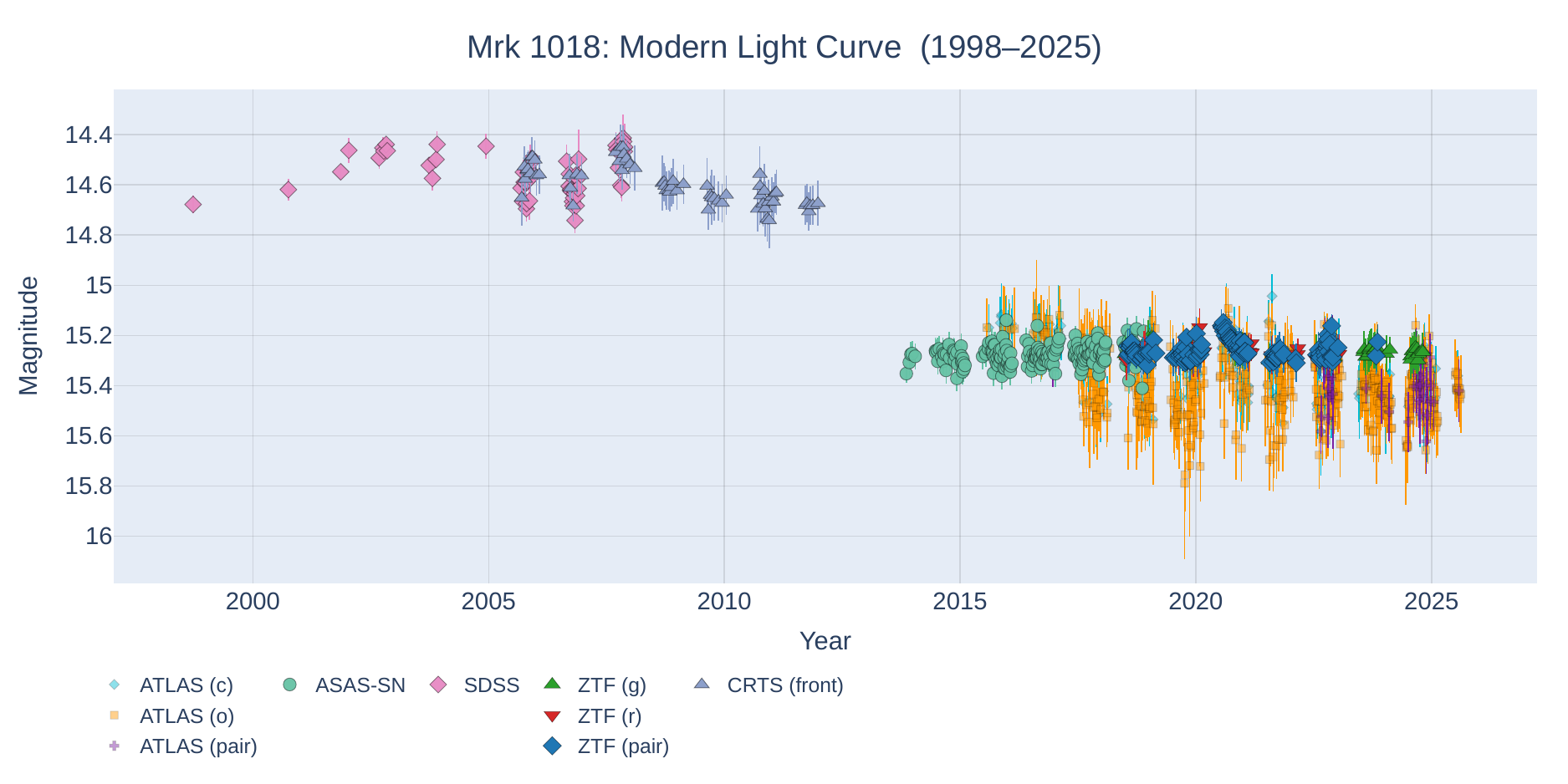}
    \caption{Mrk 1018 AGN + host modern light curve (Johnson B). The CL event can be photometrically discerned by the drop in brightness between SDSS/CRTS and ASAS-SN/ATLAS/ZTF around 2012. ATLAS and ZTF include additional host light contribution calculated from ASAS-SN measurements. All data is binned into 3 day epochs and cleaned with two iterations of 3$\sigma$ clipping.}
    \label{fig:mrk1018_b_modern}
\end{figure}

To create the modern era of Mrk 1018's light curve, data was collected from a variety of publicly available catalogues and surveys. Namely, Sloan Digital Sky Survey's Stripe 82 (SDSS, \cite{frieman_sloan_2008}), Catalina Real-time Transient Survey (CRTS, \cite{drake_first_2009}), the Zwicky Transient Facility (ZTF, \cite{bellm_zwicky_2019, graham_zwicky_2019}), the Asteroid Terrestrial-impact Last Alert System (ATLAS, \cite{tonry_atlas_2018}) and the All-Sky Automated Survey for Supernovae (ASAS-SN, \cite{shappee_all_2014, kochanek_all-sky_2017}). This data spans a 27 year baseline from 1998--2025 (see Fig. \ref{fig:mrk1018_b_modern} within \hyperref[sec:appendix_A]{Appendix A} for the enlarged light curve). Where difference images or forced photometry is available (ZTF, ATLAS \& ASAS-SN), we use the reduced images including AGN and host flux for the fair comparison between the historic data.

\section{Analysis}
\label{sec:analysis}
\subsection{Type 1 era}
One of the primary concerns when constructing the combined light curve of historic and modern data for Mrk 1018 is the differing magnitude systems across telescopes and catalogues. Empirical relations which convert from one magnitude system to another depend upon the colour of a source, and given the majority of DASCH plates are blue or unfiltered (\cite{williams_dasch_2025}), we use Mrk 1018's modern colours to make the conversion into the historic system of Johnson B band. Additionally, for CL AGN colour is seen to vary along with changing spectral type with a bluer-when-brighter trend which must be considered while treating the modern light curve which spans a transition (\cite{yang_discovery_2018}).

We perform the SDSS photometry via SExtractor on the standard SDSS filtered $g$ and $r$ frames (\cite{fukugita_sloan_1996}), using the isophotal measurement for consistency with the DASCH results. These measurements define the early time colour ($g-r$), from 1998--2007, of Mrk 1018's AGN + host when in a Type 1 state. These ranged from 0.578 to 0.714, with a median $g-r$ colour of 0.632. To convert to Johnson B band, we use the empirical relation for galaxies, ideal for Mrk 1018 as an extended source, derived from \textcite{cook_empirical_2014} Table 3:
\begin{equation} \label{eq:B}
B\approx g+0.268(g-r)+0.218
\end{equation}

CRTS photometry, spanning 2005--2011, was originally unfiltered and calibrated to Johnson V-band. CRTS uses aperture based magnitudes, which we assume includes host-flux. First we improve the photometric accuracy using the equations described in the Catalina Surveys Data Release 2 (derived using \cite{landolt_ubvri_2007, landolt_ubvri_2009}),
\begin{equation} \label{eq:landolt}
V = V_{CRTS} + 0.31(B-V)^2 + 0.04 \quad (\sigma=0.059)
\end{equation}
Here we calculate the colour index, \((B-V)\), using SDSS $(g-r)$ measurements within the same epoch (10 days), or else use the median SDSS $(g-r)$, converted using the similarly derived Cook empirical transformation for galaxies (\cite{cook_empirical_2014}): 
\begin{equation} \label{eq:cook}
(B-V)\approx 0.893(g-r)+0.161
\end{equation}
The improved photometry V-band magnitude is then converted to B-band with the addition of the $(B-V)$ colour index.
%\begin{equation} \label{eq:johnsonV}
%B = V+(B-V)
%\end{equation}

We observed an offset of the CRTS magnitudes from the Stripe 82 photometry for Mrk 1018 by $+0.135$ mag. This is within a small residual's ($+0.084\ \pm0.194$) median absolute deviation for $N=5675$ matched sources within a 7-day window epoch in Stripe 82 frames. This likely stems from the unfiltered origins of CRTS, converted into Johnson V using G0-G8 comparison stars \cite{drake_first_2009}, any sources with differing SED will introduce the $\sim$ 0.1--0.2 magnitude residuals seen. To account for this, the median offset (0.135) mag is added to the corrected CRTS magnitudes. 

% We add a third survey for Mrk 1018's Type 1 flux to verify this conclusion. The PTF photometry catalogue for Mrk 1018 includes ten SExtractor measurements (between 2009--2010) in its natural Mould-R system. We use the isophotal magnitude again for consistency with the historic data. Approximating Mould-R to the SDSS r band, we calculate the Johnson B from the Stripe 82 median $(g-r)$ and Eq. \ref{eq:B}.

\subsection{Type 2 era}
A new colour term is calculated for the appropriate conversions during Mrk 1018's fainter state using ZTF (2018--2024) \(g\) and \(r\) band observations. Observations taken within a three day period are used in tandem to calculate the colour index per epoch. Similar to the SDSS colour, the median ZTF colour ($g-r=0.881$) is used for the any other conversions made outside of its epoch. ZTF measurements themselves are converted to Johnson B using Eq. \ref{eq:B}.  

The ASAS-SN aperture photometry (16'' radius) which spans 2013--2018, is in Johnson V-band and is converted using the same method as CRTS. A similar approach is taken with ATLAS (2015--2025) whose cyan (c) and orange (o) bands are first converted to \(g-r\) with the transformations within \textcite{tonry_atlas_2018}.

\begin{figure*}[h]
    \centering
    \includegraphics[width=0.95\linewidth]{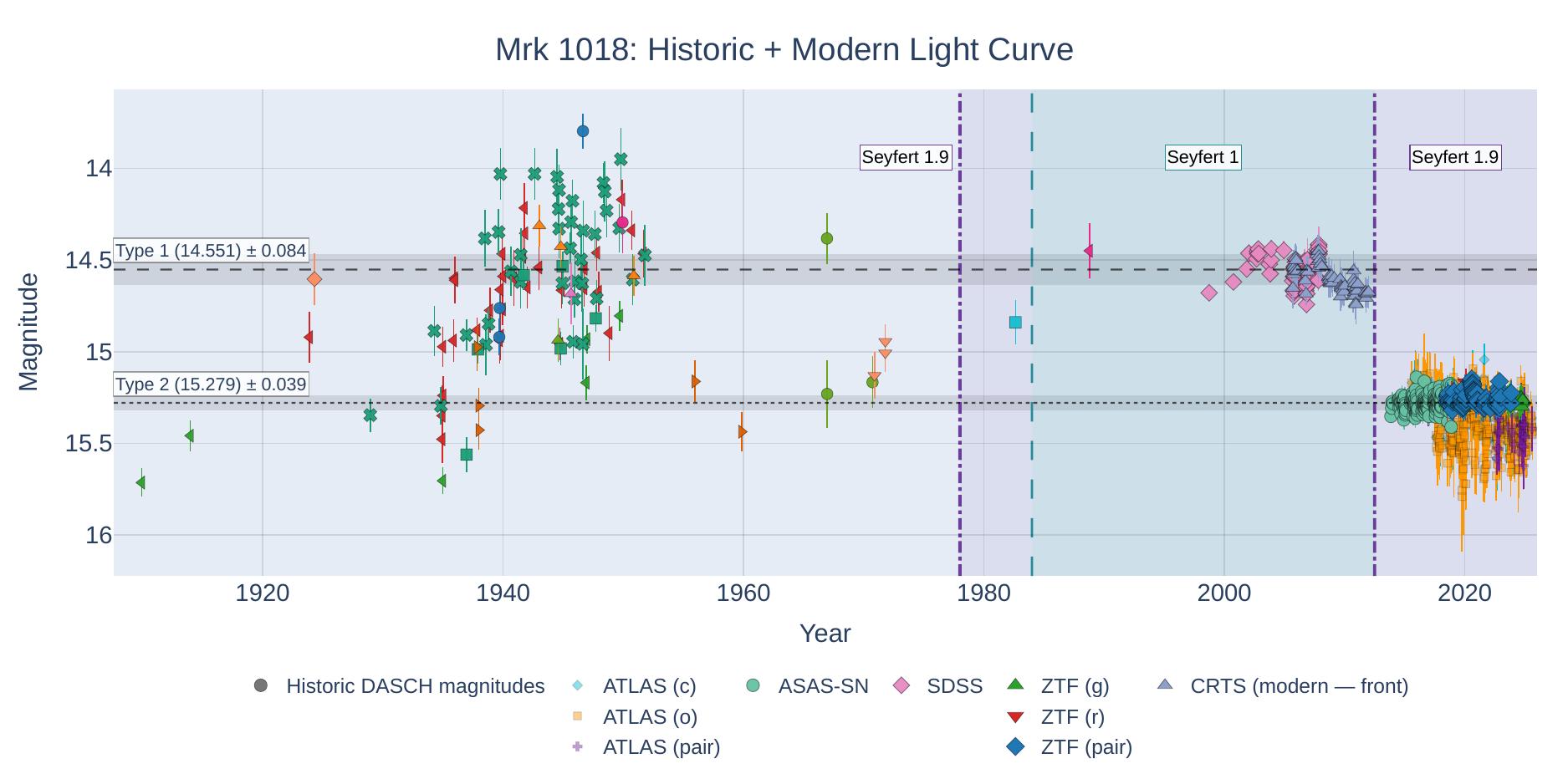}
    \caption{Mrk 1018 AGN + host 100 year Johnson B band light curve with DASCH photographic data and modern photometry (SDSS, CRTS, ASAS-SN, ATLAS and ZTF). Marker colour and shape differentiates historic data per telescope. Horizontal dashed lines and grey bands indicate Type 1 (SDSS) and Type 2 (ASAS-SN) mean magnitudes and 1 $\sigma$ ranges. Vertical dashed lines and shaded regions show spectroscopic CL events in Mrk 1018's past.}
    \label{fig:mrk1018_B}
\end{figure*}

\begin{figure*}[h]
    \centering
    \includegraphics[width=0.95\linewidth]{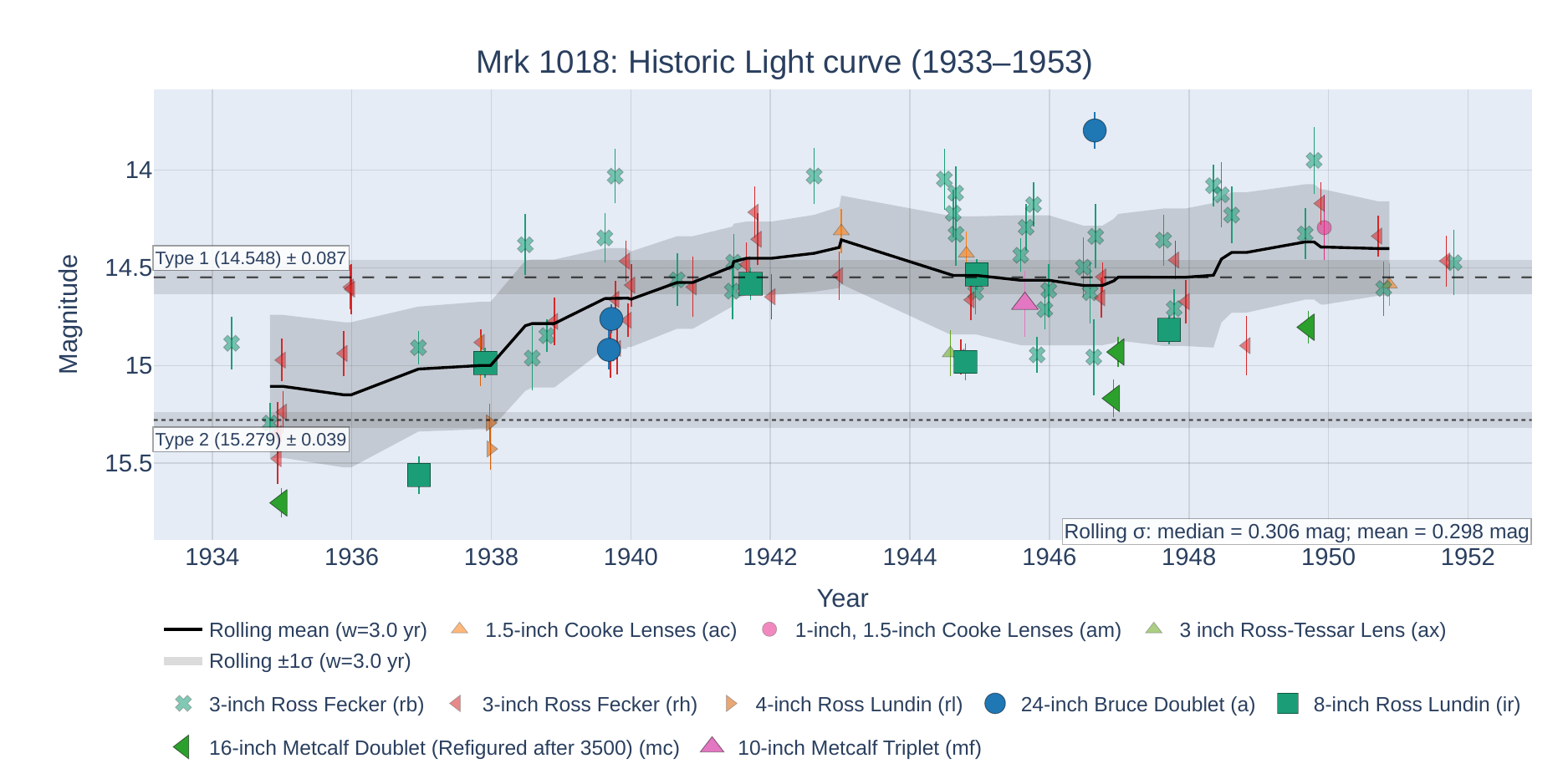}
    \caption{Mrk 1018's extreme event peaking around 1944. AGN + host flux contribution from historic DASCH Johnson B band magnitudes. The boxcar rolling mean (3 years width) is shown by the solid black line, with a 1$\sigma$ grey band. Plates taken by telescopes larger than 8 inches have larger markers, whereas smaller telescopes are indicated with 55\% opacity. Horizontal lines indicate Type 1 magnitude (SDSS; 2000--2010) and Type 2 magnitude (ASAS-SN; 2013--2018) with 1$\sigma$ bands.}
    \label{fig:historic_only}
\end{figure*}

ATLAS and ZTF use a PSF for photometry, removing the extended galaxy flux from Mrk 1018's host from the magnitudes. However, the historic and other modern magnitudes include the host's contribution to flux. Hence, we determine the host-flux by fitting it as the average difference between the median ASAS-SN flux and the median PSF flux (ATLAS and ZTF). This singular value for host flux is combined with the PSF survey's flux measurements. Losses from total galactic flux due to ASAS-SN's 16'' aperture size are likely to the order of the flux losses within the photographic plates and hence neglected.

The combined historic and modern light curve is shown in Fig. \ref{fig:mrk1018_B}. Horizontal lines indicate the brightness in the Type 1 state in the 2000's, based on the mean SDSS magnitude, and Type 2\footnote{We are aware Mrk 1018's faint state is a mostly optical Type 1.9 Seyfert, however, for simplicity, we refer to the optical type of Mrk 1018 as `Type 2' throughout the paper.} brightness from the ASAS-SN mean. A larger spread is seen in the historic data compared to the modern observations, with one $\sigma$ corresponding to 0.3 mags on a typical epoch. Given the 0.1 mag photometric accuracy of DASCH, which is mirrored by the standard deviation of the reference stars in Fig. \ref{fig:mrk1018_with_refs}, the remaining deviation is attributed to intrinsic AGN variability associated with shorter timescales which are difficult to visually discern on the 100-year light-curve. Figure \ref{fig:historic_only} better shows this short timescale variability, in addition to the significant event around 1940, with a black rolling average line and 1$\sigma$ error envelope. Larger markers highlight plates taken with telescopes greater than 8-inches in size. Data from smaller telescopes, indicated by the 55\% opaque markers, follows the same trend as the larger telescope data. While the observations are limited by the telescopes of the time and the quality of the photographic plates, the detections of Mrk 1018, which were visually confirmed as resolved sources free from defects or blending, yield a $\sigma = $ 0.3 magnitude, indicating a relatively good agreement between all the smaller telescopes.

Early observations within the 1910s indicate a low state luminosity, after which the state is not clear from the limited observations until $\sim$1930 where it appears still low. Starting from 1936, Mrk 1018's flux increases until $\sim$1943, when it settles to an increase of $19.3\ \pm 25.1 \%$ compared to the 2000s Type 1 flux value. It remains at this magnitude (average $14.50\ \pm 0.31$) until at least 1951, when the majority of observations ends. With two observations, Fig. \ref{fig:mrk1018_B} hints at a decrease back to a Type 2 luminosity around 1960. Mrk 1018 must return to a Type 2 before 1978 (when first classified; \cite{osterbrock_seyfert_1981}). The modern photometry perhaps indicates the start of Mrk 1018's transition back to a Type 1.9 Seyfert around 2007, with a slight drop in flux to 2011, although there exists a sharp decline between 2011--2013 not covered by the surveys here, to match the observed magnitudes in ASAS-SN. Hence, if this rapid drop in luminosity is the turn-off event, this work constrains a new maximum turn-off duration of 683 days ($\sim1.9$ yr), with previous estimates between 2--3 years (\cite{saha_close_2025}). This supports CL mechanisms which can drastically affect the brightness of the source within particularly short timescales, such as via a changing mode of accretion. 

We calculate a possible period using a Generalised Lomb-Scargle (GLS) periodogram of the historic and modern data. We find a broad GLS maximum, peaking at $P=47.32\ \pm 0.73$ years. The GLS normalised power against log-frequency can be found in \hyperref[sec:appendix_A]{Appendix A}. We test this result with bootstrap resampling ($N=200$), yielding a median of $46.96$ yr, and IQR 46.31--47.77. A jackknife split gives a $P_{Early}=49.64$ yr and $P_{Late}=50.09$ yr to confirm neither modern or historic bias. A shuffle-based null test yields a 95th percentile maximum LS power of 172.39 around the candidate period (200 shuffles), well below the observed LS power of 9845.75. From this we calculate an empirical false alarm probability of 0.005.

\section{Discussion}
\label{sec:discussion}
% Event discussion
From the light curve, Mrk 1018's historic brightness from 1943--1951 is consistent with its Type 1 brightness seen in the 2000s. Although similar extreme AGN variability (in excess of one magnitude) is not often associated with CL events (on the order of 20\%; \cite{macleod_changing-look_2019}), given Mrk 1018's history as a recurrent CL AGN, it is compelling to attribute such dramatic change in brightness to an additional changing look event.

% Period discussion
The period derived from the peak of the power spectrum, $47.32\ \pm 0.73$ years, does not fall within the expected range from the previous transitions, 28--34 years. However, the GLS spectrum shows a broad low frequency plateau, where the first major peak is at $P\sim29$ yr, which is more consistent with the expectation. This likely tells us that there is a characteristic several-decade timescale associated with Mrk 1018's variability rather than a strictly periodic cycle. There are two further inconsistencies with a truly periodic mechanism: the transition durations and the Type 1 phase duration. The historic light curve indicates an initial brightening duration of $\sim$7 years, and a potential fading duration no longer than 9 years. This is consistent with a typical changing look event timescale (\cite{ricci_changing-look_2023}), although somewhat longer than Mrk 1018's previously recorded transitions with a maximum 6 years turn-on (between 1978 and 1984; \cite{osterbrock_seyfert_1981, cohen_variability_1986}), and less than $\sim$1.9 years turn-off (2011--2013). Additionally, if Mrk 1018 turns-off around 1960, this historic Type 1 phase lasts no longer than $\sim$20 years. This is notably shorter than than the 28--34 year duration of the previously observed bright phase ($\sim$1984--2012). Both of these inconsistencies disfavour a strictly periodic CL mechanism in Mrk 1018.

% Implications 
Our data show (i) century-scale variability with extended low states and asymmetric bright phases, (ii) a dominant and broad low-frequency feature at $\sim$29--47 yr that remains stable under our colour conversions, and (iii) no persistent narrow peak above noise/aliases in our period analyses. These findings are consistent with expectations from Chaotic Cold Accretion (CCA; e.g., \cite{gaspari_raining_2017,veronese_interpreting_2024}), where turbulent condensation and cloud–cloud collisions intermittently enhance inflow, yielding stochastic variability with a characteristic correlation timescale (e.g.~eddy turnover time at micro scales) rather than a strictly periodic signal. In this framework, the 29–-47 yr feature is interpreted as a characteristic timescale (not a coherent oscillation), and the 1935-–1955 episode (20-–25 yr) as a single extended event. CCA therefore offers a self-regulated, stochastic AGN feeding explanation for the long-term modulation, without invoking a mode-locked cycle. Although, given only two full periods and gaps within the 100-year light curve, an alternative consideration is that this characteristic period may be an artefact of intrinsic AGN red noise \cite{krishnan_detection_2021} triggering the changing modes of accretion. More observations are required to better understand the nature of Mrk 1018's Changing Look. In line with this, if Mrk 1018's mechanism is a result of a characteristic correlation timescale as opposed to intrinsic noise, we predict a turn-on transition for Mrk 1018 to begin between 2033 and 2045, based on previous CL activity (1935, 1955--1960, 1978--1984, 2011--2013).

\section{Conclusions}
\label{sec:conclusion}
The historic DASCH light curves combined with modern data show that Mrk 1018 underwent a significant brightening event from $\sim$1936--1950 consistent with its Type 1 luminosity seen in the 2000s. This is a new CL event in Mrk 1018's past and shows that such events recur. An analysis of the entire photometric behaviour, suggests a characteristic timescale between 29--47 years, indicative of a long-term modulation of spectral type rather than a strictly periodic CL mechanism. Additionally, a new turn-off duration is constrained for the 2012 event of $\sim1.9$ years, supporting a CL mechanism for Mrk 1018 with the capability for drastic luminosity drops in short timescales.

These archived and digitised plates are a unique opportunity to peer into the past and examine AGN in an unprecedented way. Yet there is much more to be done; analysing other known CL AGN for previous, perhaps periodic events, or searching the historic data for unknown CL events to identify objects of interest for upcoming surveys such as LSST (\cite{ivezic_lsst_2019}). With these next-generation time domain surveys coming online, historic records hold a wealth of knowledge that we may use to determine what we can expect from such surveys.

\section*{Data Availability}
All historic photometry is available through the DASCH python package (\href{https://daschlab.readthedocs.io/}{\textit{daschlab}}). The modern light curve data was accessed through the CRTS, ASAS-SN, ATLAS and ZTF public catalogues. Stripe 82 frames were accessed through the SDSS Data Archive Server.

\section*{Acknowledgements}
We acknowledge and pay respect to the traditional owners of the land on which our universities are situated, upon whose unceded, sovereign, ancestral lands we work. We pay respects to their Ancestors and descendants, who continue cultural and spiritual connections to Country.

M.K. is supported by the DLR grant FKZ 50 OR 2519.
M.G. acknowledges support from the ERC Consolidator Grant \textit{BlackHoleWeather} (101086804).

\printbibliography
\onecolumn

\section*{Apendix A: Additional Figures}
\label{sec:appendix_A}

\begin{figure}[h]
    \centering
    \includegraphics[width=\textwidth]{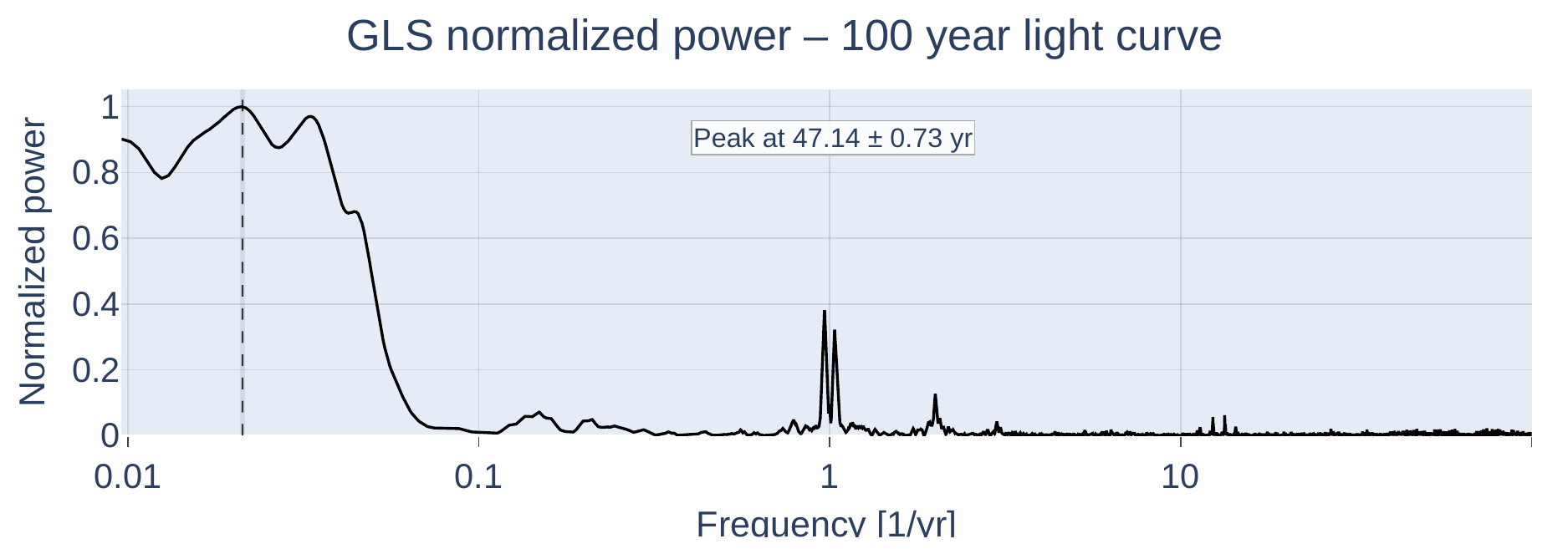}
    \caption{Combined GLS for Mrk 1018 with log-frequency in 1/yr on the x-axis and normalised LS power on the y-axis. The peak at 29.31 $\pm\ 0.18$ years is marked by vertical line, and is found within a broad shoulder, indicating a characteristic timescale, rather than strict period. The minor peaks at 1 year are due to the seasonal alias.}
    \label{fig:LS}
\end{figure}

\FloatBarrier

\end{document}